\documentclass[a4paper]{article}

\usepackage{INTERSPEECH2020}
\title{A Pyramid Recurrent Network for Predicting Crowdsourced Speech-Quality Ratings of Real-World Signals}
\name{Xuan Dong and Donald S. Williamson\thanks{This research was supported by a NSF grant (IIS-1755844).}}
\address{Department of Computer Science, Indiana University, USA}
\email{xuandong@iu.edu, williads@indiana.edu}

\begin{document}

\maketitle
\begin{abstract}
The real-world capabilities of objective speech quality measures are limited since current measures (1) are developed from simulated data that does not adequately model real environments; or they (2) predict objective scores that are not always strongly correlated with subjective ratings. Additionally, a large dataset of real-world signals with listener quality ratings does not currently exist, which would help facilitate real-world assessment. In this paper, we collect and predict the perceptual quality of real-world speech signals that are evaluated by human listeners. We first collect a large quality rating dataset by conducting crowdsourced listening studies on two real-world corpora. We further develop a novel approach that predicts human quality ratings using a pyramid bidirectional long short term memory (pBLSTM) network with an attention mechanism. The results show that the proposed model achieves statistically lower estimation errors than prior assessment approaches, where the predicted scores strongly correlate with human judgments.

\end{abstract}


\section{Introduction}

Subjective listening studies are the most reliable form of speech quality assessment for many applications, including speech enhancement and audio source separation~\cite{hu2007evaluation, emiya2011subjective}. Listeners often rate the perceptual quality of testing materials using categorical or multi-stimuli rating protocols~\cite{rec2008p, series2014method}. The test materials are often artificially created by additively or convolutionally mixing clean speech with noise or reverberation at prescribed levels, to simulate real environments~\cite{hirsch2000aurora, kinoshita2016summary}. Unfortunately, the simulated data does not capture all the intricate details of real environments (e.g.~speaker and environmental characteristics), so it is not clear if these assessments are consistent with assessment results from real-world environments. Many investigations conclude that more realistic datasets and scenarios are needed to improve real-world speech processing performance~\cite{mclaren2016speakers, reddy2020interspeech, barker2018fifth}. However, the cost and time-consuming nature of subjective studies also hinders progress.

Computational objective measures enable low cost and efficient speech quality assessment, where many intrusive, non-intrusive, and data-driven approaches have been developed. Intrusive measures, such as the perceptual evaluation of speech quality (PESQ)~\cite{pesq2001}, signal-to-distortion ratio (SDR)~\cite{emiya2011subjective} and perceptual objective listening quality analysis (POLQA)~\cite{beerends2013perceptual}, generate quality scores by calculating the dissimilarities between a clean reference speech signal and its degraded counterpart (e.g.~noisy, reverberant, enhanced). These measures, however, do not always correlate well with subjective quality results~\cite{cano2016evaluation, santos2014improved}. 

Several non-intrusive (or reference-less) objective quality measures have been developed, including the ITU-T standard P.563~\cite{malfait2006p}, ANSI standard ANIQUE+~\cite{kim2007anique+}, and the speech to reverberation modulation energy ratio (SRMR)~\cite{falk2010non}. These approaches use signal processing concepts to generate quality-assessment scores. These approaches, however, rely on signal properties and assumptions that are not always realized in real-world environments, hence the assessment scores are not always consistent with human ratings~\cite{kinoshita2016summary, mittag2019non}. More recent work uses data-driven methods to estimate speech quality~\cite{fu2018quality, mittag2019non, avila2019non, dong2019classification, dong2020attention}. The authors in \cite{sharma2016data} combine hand-crafted feature extraction with a tree-based regression model to predict objective PESQ scores. Quality-Net~\cite{fu2018quality} provides frame-level quality assessment by predicting the utterance-level PESQ scores that are copied as per-frame labels using a bidirectional long short-term memory (BLSTM) network. Similarly, NISQA~\cite{mittag2019non} estimates the per-frame POLQA scores using a convolutional neural network (CNN). It subsequently uses a BLSTM to aggregate frame-level predictions into utterance-level objective quality scores. These data-driven approaches perform well and increase the practicality of real-world assessment. However, the usage of objective quality scores as training targets is a major limitation, since objective measures only approximate human perception~\cite{emiya2011subjective, cano2016evaluation}. Alternatively, the model developed in \cite{avila2019non} predicts the mean opinion score (MOS)~\cite{rec2006p} of human ratings, but the ratings are collected on simulated speech data. This approach advances the field, but it is not enough to ensure good performance in real environments. A complete approach is needed that predicts human quality ratings of real recordings.

In this study, we conduct a large-scale listening test on real-world data and collect 180,000 subjective quality ratings through Amazon's Mechanical Turk (MTurk)~\cite{paolacci2010running} using two publically-available speech corpora~\cite{stupakov2009cosine, richey2018voices}. This platform provides a diverse population of participants at a significantly lower cost to facilitate accurate and rapid testing~\cite{ribeiro2011crowdmos, schoeffler2015towards, cartwright2016fast}. These corpora have a wide range of distortions that occur in everyday life, which reflect varying levels of noise and reverberation. Our listening tests follow the MUltiple Stimuli with Hidden Reference and Anchor (MUSHRA) protocol~\cite{series2014method}. To the best of our knowledge, a large publically-available dataset that contains degraded speech and human quality ratings does not currently exist. We additionally develop an encoder-decoder model with attention mechanism~\cite{chorowski2015attention} to non-intrusively predict the perceived speech quality of these real-world signals. The encoder consists of stacked pyramid BLSTMs~\cite{chan2016listen} that convert low-level speech spectra into high-level features. This encoder-decoder architecture reduces the sequential size of the latent representation that is provided to an attention model. The key difference between this proposed approach and related approaches, is that our approach predicts mean-opinion scores of real-world signals using a novel deep-learning framework. The following sections discuss the details and results of our approach.

\section{Methods}
\label{sec:method}

\subsection{Crowdsourced listening study procedures}

We create human intelligence tasks (HIT) on Amazon Mechanical Turk (MTurk) for our crowdsourced subjective listening test~\cite{rec2018crowdspeech}, where each HIT is completed by 5 crowdworkers (i.e. subjective listeners). At the beginning of each HIT, crowdworkers are presented with instructions that describe the study's purpose and procedures. The study has a qualification phase that collects demographic information (e.g.~age group, gender, etc.). We also collect information about their listening environment and devices they are using to hear the signals. The participants are required to be over 18 years of age, native English speakers, and have normal hearing. This study has been approved by Indiana University's Institutional Review Board (IRB). A small monetary incentive was provided to all approved participants.

Each HIT contains 15 trials of evaluations that follow the recommendation of ITU-R BS.1534 (a.k.a.~MUSHRA)~\cite{series2014method}. Each trial has multiple stimuli from varying conditions including a hidden clean reference, an anchor signal (low-pass filtered version of the clean reference) and multiple real-world noisy or reverberant speech signals (i.e., test stimuli). After listening to each signal, the participants are asked to rate the quality of each sound on a continuous scale from 0 to 100 using a set of sliders. We clarify the quality scale, so that sounds with excellent quality should be rated high (i.e., 81 $\sim$ 100) and bad quality sounds should be rated low (i.e., 1 $\sim$ 20). The listener is able to play each stimuli as often as desired. Each HIT typically takes 12 minutes or less to complete.

Overall, we launched 700 HITs. 3,578 crowdworkers participated in our listening tests, and 3,500 submissions were approved for subsequent usage. 2,045 crowdworkers are male and 1,455 are female. Their ages cover a range from 18 to 65. 2,837 of them have prior experience with listening tests.

\subsection{Speech material}

Previous listening studies use artificially created noisy- or reverberant-speech stimuli ~\cite{emiya2011subjective, naderi2015effect, cartwright2016fast, avila2019non}. This enables control over the training and testing conditions, however, it limits external validity as the designed distortions differ from those in real environments. Therefore, we use two speech corpora that were recorded in a wide range of real environments.

We first use the COnversational Speech In Noisy Environments (COSINE) corpus~\cite{stupakov2009cosine}. This corpora contains 150 hours of audio recordings that are captured using 7-channel wearable microphones, including a close-talking mic (near the mouth), along with shoulder and chest microphones. It contains multi-party conversations about everyday topics in a variety of noisy environments (such as city streets, cafeterias, on a city bus, wind noise, etc). The audio from the close-talking microphone captures high quality speech and is used as the clean reference. Audio from the shoulder and chest microphones capture significant amounts of background noise and speech, hence they serve as the noisy signals under test. For each close-talking signal, one noisy signal (from shoulder or chest) is used alongside the reference and anchor signals, and evaluated by the listeners using the MUSHRA procedure. The approximated signal-to-noise ratios (SNRs) of the noisy signals range from -10.1 to 11.4 dB.

We also use the Voices Obscured in Complex Environmental Settings (VOiCES) corpus~\cite{richey2018voices}. VOiCES was recorded using twelve microphones placed throughout two rooms of different sizes. Different background noise are played separately in conjunction with foreground clean speech, so the signals contain noise and reverberation. The foreground speech is used as the reference signal, and the audio captured from two of the microphones are used as reverberant stimuli. The approximated speech-to-reverberation ratios (SRRs) of these signals range from -4.9 to 4.3 dB.

In the listening tests, we deploy 18,000 COSINE signals and 18,000 VOiCES signals. Each stimulus is truncated to be 3 to 6 seconds long. In total 45 hours of speech signals are generated and 180k subjective human judgments are collected. 

\subsection{Data cleaning and MOS calculation}

A crowdworker's responses are rejected if the response contains malicious behavior~\cite{gadiraju2015understanding}, such as random scoring or the amount of unanswered responses exceeds 20\% of the HIT. Data cleaning is then performed to remove rating biases and obvious outliers. Some participants tend to rate high, while others tend to rate low. This potentially presents a challenge when trying to predict opinion scores~\cite{zielinski2008some}. The following steps alleviate this problem. 

The Z-score of each stimuli is first calculated across each condition. Responses with absolute Z-scores above 2.5 are identified as potential outliers~\cite{han2011data}. The ratings of all unanswered trials are removed in this step as well. A rescaling step is then performed to normalize the rating ranges amongst all crowdworkers. Specifically, Min-max normalization is performed, and the new rating scale is from 0 to 10. 

A consensus among crowdworkers is expected over the same evaluated stimulus. If the rating of one crowdworker has very low agreement with the other crowdworkers, this rating is considered inaccurate or a random data point. Thus, we apply two robust non-parametric techniques, density based spatial clustering of applications with noise (DBSCAN)~\cite{ester1996density} and isolation forests (IF)~\cite{liu2008isolation}, to discover outliers that deviate significantly from the majority ratings. DBSCAN and IF are used in an ensemble way, and a conservative decision is made in which ratings are only discarded when both algorithms identify it as an outlier. The algorithms are implemented with scikit-learn using default parameters.

\begin{figure}[t]
	\centering
	\minipage{0.25\textwidth}
	\includegraphics[width=\linewidth]{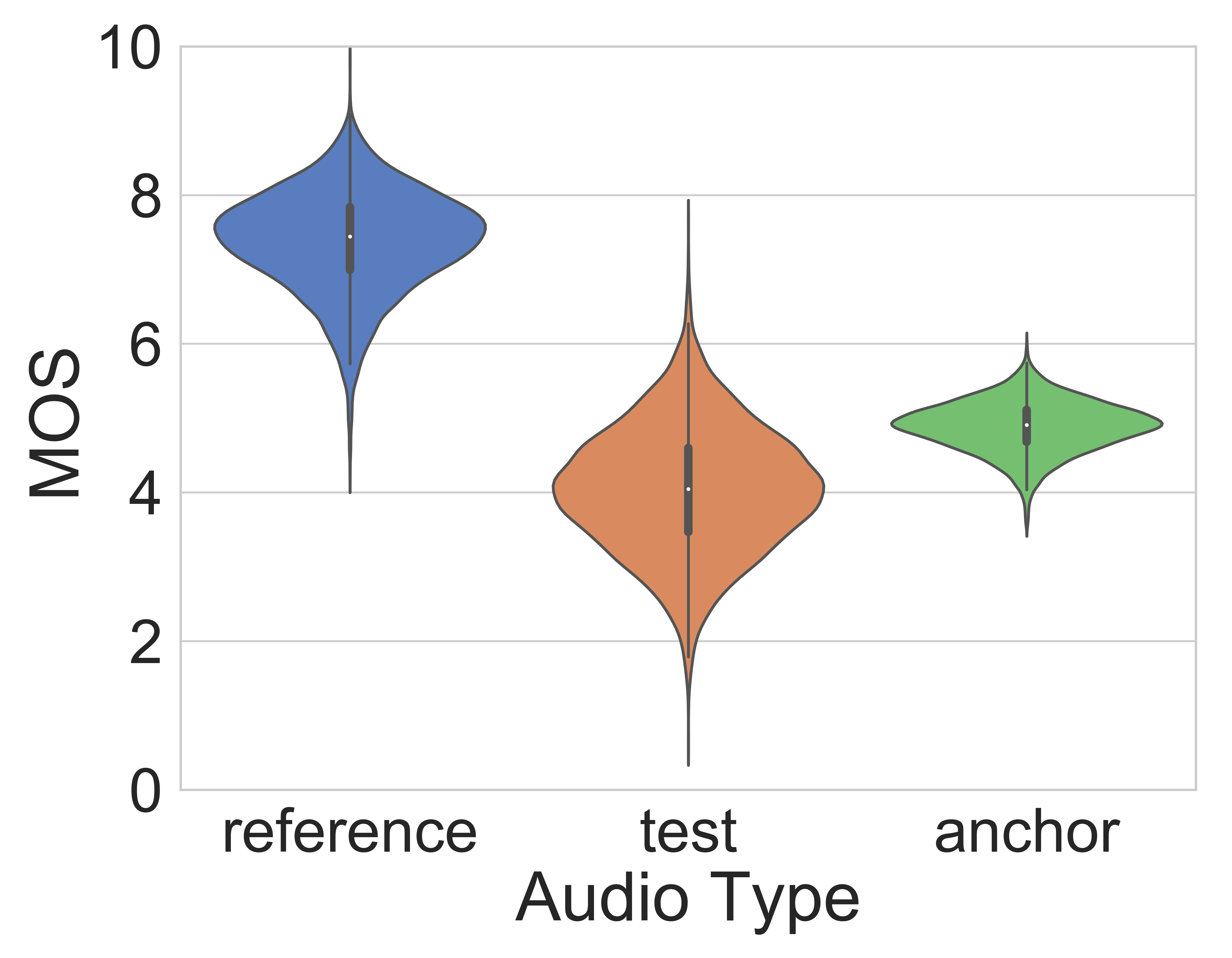}
	\endminipage
	\minipage{0.25\textwidth}
	\includegraphics[width=\linewidth]{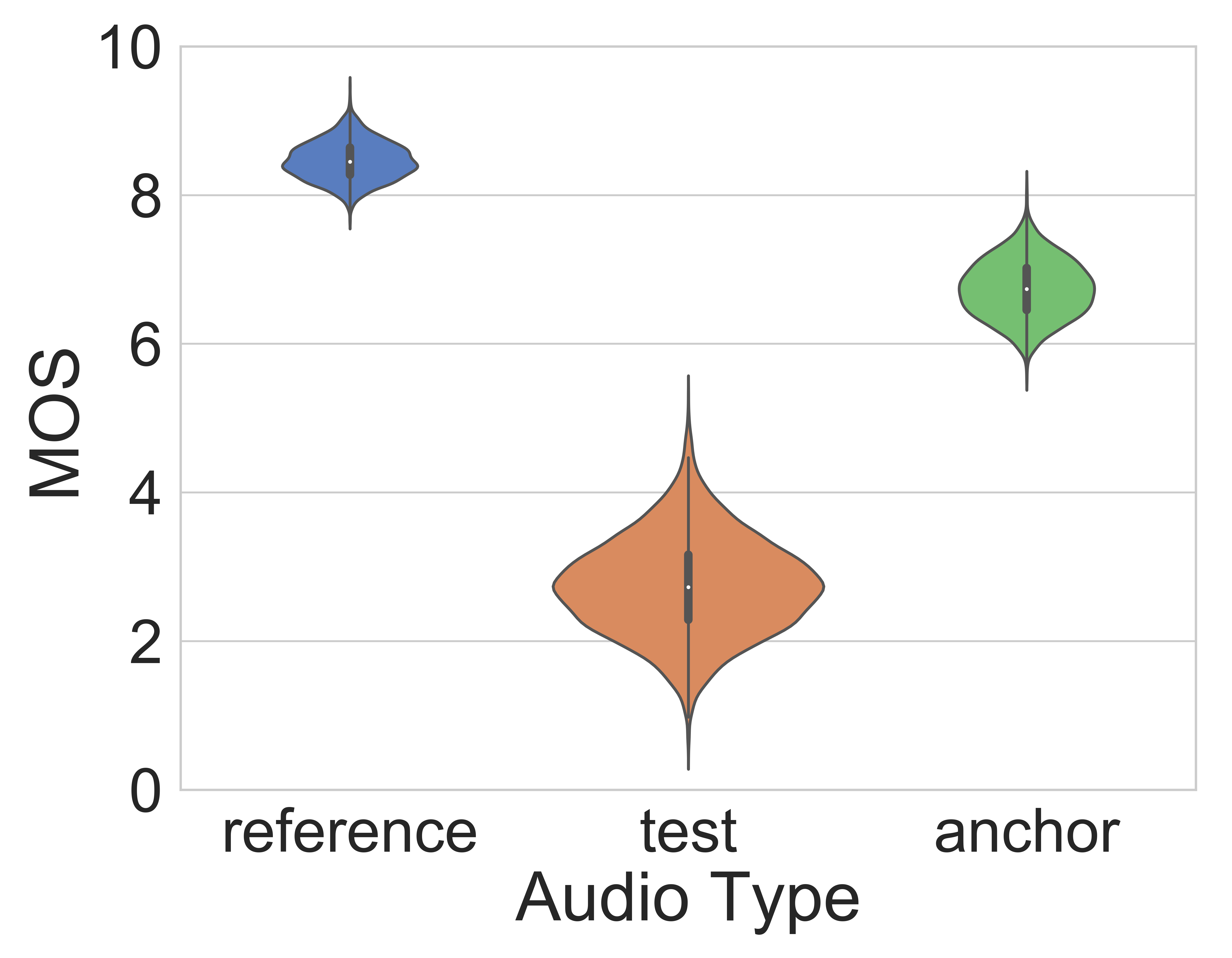}
	\endminipage
	\caption{MOS distributions of COSINE (left) and VOiCES (right) corpora.}
	\label{fig:mosboxplots}
\end{figure}

After data cleaning is complete, the scaled ratings for each stimulus are averaged and this is used as the MOS for the corresponding signal. The full distribution of the scaled MOS of each speech corpus is shown in Figure~\ref{fig:mosboxplots}. As expected, the reference signals are rated high and the anchor signals have a relatively narrow range. The test stimuli of COSINE data varies from 2.0 to 6.0, while those from VOiCES are concentrated between 1.5 to 4.0. Major outliers seldomly occur in each condition. 

\subsection{Data-driven MOS quality prediction}

\begin{figure}[t]
	\centering
	\includegraphics[width=0.95\linewidth]{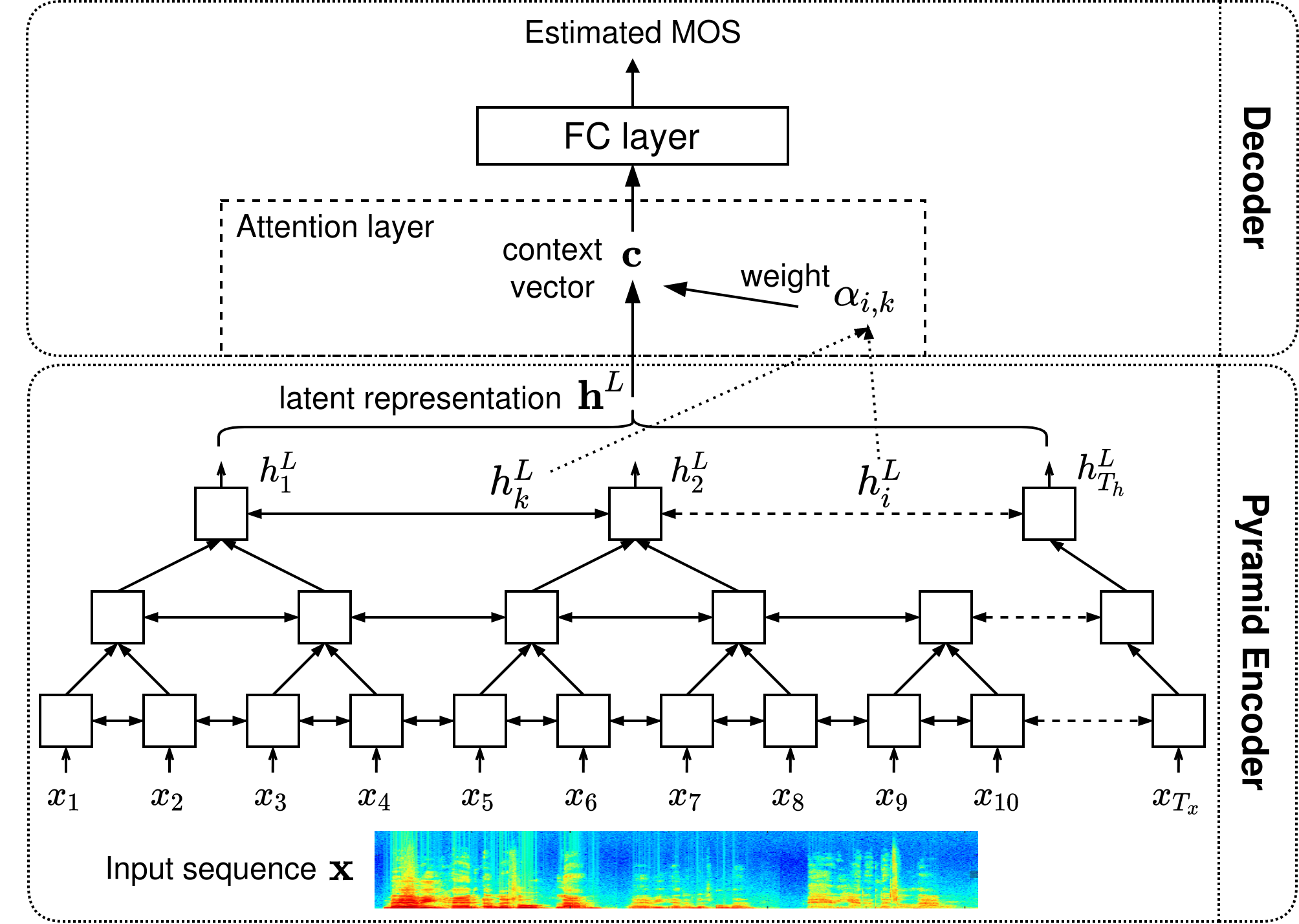}
	\caption{Illustration of the proposed attention-based pyramid BLSTM model for predicting MOS scores. Only two pBLSTM layers are displayed.}\label{fig:arch_pBLSTM}
\end{figure}

Our proposed attention-based encoder-decoder model for predicting human quality ratings of real-world signals is shown in Fig.~\ref{fig:arch_pBLSTM}. The approach consists of an encoder that converts low-level speech signals into higher-level representations, and a decoder that translates these higher-level latent features into an utterance-level quality score (e.g.~the predicted MOS). The decoder specifies a probability distribution over sequences of features using an attention mechanism~\cite{chorowski2015attention}. 

The encoder utilizes a stacked pyramid bidirectional long short term memory (pBLSTM)~\cite{chan2016listen} network, which has been successfully used in similar speech tasks (ASR~\cite{chan2016listen} and voice conversion~\cite{zhang2019sequence}), but not for speech assessment. Utterance-level prediction is challenging since the signals may be long, which complicates convergence and produces inferior results. The connections and layers of a pyramidal architecture enable processing of sequences at multiple time resolutions, which effectively captures short- and long-term dependencies. 

Fig.~\ref{fig:arch_pBLSTM} depicts an unrolled pBLSTM network. The boxes correspond to BLSTM nodes. The input to the network, $x_t$, is one-time frame of the input sequence at the $t$-th time step. In a pyramid structure, the lower layer outputs from $M$ consecutive time frames are concatenated and used as inputs to the next pBLSTM layer, along with the recurrent hidden states from the previous time step. More generally, the pBLSTM model for calculating the hidden state at layer $l$ and time step $t$ is
\begin{equation}
h_t^l = \mathrm{pBLSTM} \left( h_{t-1}^l, \mathrm{Concat}(h_{M*t-M+1}^{l-1}, \dots, h_{M*t}^{l-1}) \right),
\end{equation}
where $M$ is the reduction factor between successive pBLSTM layers. In the implementation, we use $L=3$ pBLSTM layers (with 128, 64 and 32 nodes in each direction, respectively) on
top of a BLSTM layer that has 256 nodes. The factor $M=2$ is adopted here, same as \cite{chan2016listen}. This structure reduces the time resolution from the input $\mathbf{x}$ to the final latent representation $\mathbf{h^L}$ by a factor of $M^3 = 8$. The encoder output is generated by concatenating the hidden states of the last pBLSTM layer into vector $\mathbf{h^L}$ (e.g  $\{h_1^L, h_2^L, \dots, h_{T_h}^L \}$). $T_h$ is the number of final hidden states.

The decoder is implemented as an attention layer followed by a fully-connected (FC) layer. The self-attention mechanism~\cite{vaswani2017attention} uses the encoder output at the $i$-th and $k$-th time steps (e.g.,~$h_i^L$ and $h_k^L$) to compute the attention weights: $\alpha_{i,k} = \mathrm{Attention} (h_i^L, h_k^L)$. A context vector $c_i$ is computed as a weighted sum of the encoder hidden states:
$c_i = \sum_{k=1}^{T_h} \alpha_{i,k} h_k^L$. Note that the pyramid structure of the encoder results in shorter latent representations than the original input sequence, and it leads to fewer encoding states for attention calculation at the decoding stage. Finally, the context vector of the decoder is passed to a FC layer that has 32 hidden units and results in an estimate of the perceptual quality (i.e., MOS). The model is optimized using the mean-squared error loss and Adam optimization. It is trained for 100 epochs. The aforementioned parameters were empirically determined based on best performance with a validation set. 

\section{Experiments and analysis}
\label{sec:experiment}

\subsection{Experimental setup}

The speech corpora from both datasets consist of 16-bit single channel files sampled at 16 kHz. For MOS prediction, the input speech signals are segmented into 40 ms length frames, with 10 ms overlap. A FFT length of 512 samples and a Hanning window are used to compute the spectrogram. Mean and variance normalization are applied to the input feature vector (i.e., log-magnitude spectrogram). The noisy or reverberant stimuli of each dataset are divided into training (70\%), validation (10\%) and testing (20\%) sets, and trained separately. Five fold cross-validation is used to assess generalize performance to unseen data.

Four metrics are used to evaluate MOS prediction: the mean absolute error (MAE); the epsilon insensitive root mean squared error (RMSE$^{\star}$)~\cite{rec2012p}, which incorporates a 95\% confidence interval when calculating prediction errors; Pearson's correlation coefficient $\gamma$ (PCC); and Spearman's rank correlation coefficient $\rho$ (SRCC), which assesses monotonicity.

\begin{table*}[t]
	\renewcommand\thetable{II}
	\centering
	\caption{Performance comparison with the state-of-the-art non-intrusive methods on each corpus.}
	\label{tab:sota_compare}
	\resizebox{0.8\textwidth}{!}{%
		\begin{tabular}{@{}lcccccccc@{}}
			\toprule
			& \multicolumn{4}{c}{COSINE} & \multicolumn{4}{c}{VOiCES} \\ \cmidrule(l){2-5} \cmidrule(l){6-9}
			& MAE & RMSE$^{\star}$ & PCC ($\gamma$) & SRCC ($\rho$) & MAE & RMSE$^{\star}$ & PCC ($\gamma$) & SRCC ($\rho$) \\ \midrule
			P.563~\cite{malfait2006p} & 0.85 & 0.94 & 0.55 & 0.54 & 1.09 & 1.31 & -0.06 & -0.05 \\
			SRMR~\cite{falk2010non} & 1.37 & 1.81 & 0.39 & 0.43 & 0.76 & 0.92 & 0.61 & 0.62 \\
			AutoMOS~\cite{patton2016automos} & 0.74 & 0.83 & 0.75 & 0.79 & 0.75 & 0.78 & 0.76 & 0.75 \\
			Quality-Net~\cite{fu2018quality} & 0.66 & 0.70 & 0.82 & 0.85 & 0.70 & 0.72 & 0.81 & 0.82 \\
			DNN~\cite{avila2019non} & 0.57 & 0.65 & 0.85 & 0.86 & 0.73 & 0.70 & 0.86 & \textbf{0.86} \\
			NISQA~\cite{mittag2019non} & 0.53 & 0.59 & 0.89 & 0.88 & 0.68 & 0.75 & 0.84 & 0.85 \\ \midrule
			\textbf{pBLSTM + Attn} & \textbf{0.45} & \textbf{0.52} & \textbf{0.91} & \textbf{0.90} & \textbf{0.55} & \textbf{0.61} & \textbf{0.88} & \textbf{0.86} \\ \bottomrule
		\end{tabular}%
	}
\end{table*}

\begin{table}[t]
	\renewcommand\thetable{I}
	\centering
	\caption{Performance comparison with baseline models. Results on two corpora are reported together.}
	\label{tab:baseline_compare}
	\resizebox{0.45\textwidth}{!}{%
		\begin{tabular}{@{}lcccc@{}}
			\toprule
			& MAE  & RMSE$^{\star}$ & PCC ($\gamma$) & SRCC ($\rho$) \\ \midrule
			BLSTM         & 0.85 & 0.96           & 0.53           & 0.52          \\
			pBLSTM        & 0.79 & 0.92           & 0.56           & 0.56          \\
			BLSTM + Attn  & 0.68 & 0.74           & 0.80           & 0.79          \\
			pBLSTM + Attn & 0.51 & 0.57           & 0.89           & 0.88          \\ \bottomrule
		\end{tabular}%
	}
\end{table}

\subsection{Prediction of subjective quality}

The proposed model is denoted as pBLSTM+Attn, and we first compare with three baseline models. The first model replaces the pBLSTM layers with conventional BLSTM layers (denoted as BLSTM+Attn), in order to determine the benefit of the pyramid structure. All other hyper-parameters are kept unchanged. The second and third baseline models remove the attention mechanism from the proposed model and the BLSTM model, respectively, and are denoted as pBLSTM and BLSTM. These models assesses how much the attention module contributes to the overall performance. 

The results for the baseline and proposed models are presented in Table \ref{tab:baseline_compare}. It can be seen that, on average, the proposed model outperforms all baseline models according to all metrics. The pyramid architecture (pBLSTM) improves the performance of the encoder, since it captures global and local dependencies in the latent representation space. This results in average correlations of $\rho=0.89$ and $\gamma=0.88$ with pBLSTM+Attn, which are much higher than the $\rho=0.53$ and $\gamma=0.52$ with BLSTM, and $\rho=0.80$ and $\gamma=0.79$ with BLSTM+Attn model. The influence of attention is observed by comparing BLSTM or pBLSTM performance with their attention counterparts. For instance, the RMSE$^{\star}$ drops from 0.96 for the BLSTM to 0.74 for the BLSTM+Att. pBLSTM+Attn reduces the MAE from 0.79 to 0.51 and increases the PCC from 0.56 to 0.89, due to the incorporation of an attention layer. These results further confirm the effectiveness of the attention module. A statistical significance test indicates these results are statistically significant ($p$-value $< 0.0001$). 

Next, we compare our model with six non-intrusive methods, including two conventional measures that are based on voice production and perception, and four data-driven approaches that utilize deep learning. P.563~\cite{malfait2006p} essentially detects degradations by a vocal tract model and then reconstructs a clean reference signal. SRMR~\cite{falk2010non} is an auditory-inspired model which utilizes the modulation envelopes of the speech signal to quantify speech quality. Since the output ranges of P.563 and SRMR are different from our scaled MOS (i.e., 0 to 10), a 3rd order polynomial mapping suggested by ITU P.1401~\cite{rec2012p} is used to compensate the outputs when calculating MAE and RMSE$^{\star}$. AutoMOS~\cite{patton2016automos} consists of a stack of two LSTMs and takes a log-Mel spectrogram as input. Quality-Net~\cite{fu2018quality} uses one BLSTM and two FC layers. NISQA~\cite{mittag2019non} uses a combination of six CNN and two BLSTM layers. In \cite{avila2019non}, a deep neural network (DNN) with four hidden layers is used, where it generates utterance-level MOS estimates from the frame-level predictions. Each of these approaches are trained with the same data split as the proposed model to predict the MOS scores, using the approach's default parameters.

As can be seen from the results in Table \ref{tab:sota_compare}, all data-driven approaches outperform the conventional measures (i.e., P.563 and SRMR) with a good margin. This is due, in great part, by the fact that conventional measures do rely on the assumptions that are not always true in real environments, while the data-driven approaches are able to learn informative features automatically. 

When comparing to recent data-driven approaches, the proposed model achieves the highest performance in terms of both prediction error and the correlations with the ground-truth MOS, except for SRCC of the DNN model on VOiCES data ($\gamma$ = 0.86). The proposed model, however, achieves higher correlations, $\rho=0.91$ and $\gamma=0.90$ than the $\rho=0.85$ and $\gamma=0.86$ of the DNN model on COSINE data. The PCC of the proposed model also far exceeds the 0.75 of AutoMOS and 0.82 of Quality-Net. Similar trends occur on VOiCES data as well. pBLSTM+Attn improves the PCC to 0.88, compared to AutoMOS with 0.76, Quality-Net with 0.81, and NISQA with 0.84. Additionally, the proposed pBLSTM+Attn achieves RMSE$^{\star}$ of 0.52 and 0.61 on COSINE and VOICES data, respectively, which clearly outperforms the 0.83 and 0.78 of AutoMOS, the 0.70 and 0.72 of Quality-Net, and 0.65 and 0.70 of DNN. Our MAE and RMSE$^{\star}$ scores are also lower than NISQA. Our model shows statistical significance (e.g. $p$ $< 0.01$) against all approaches and metrics, except for MAE and RMSE$^{\star}$ on the COSINE data with NISQA where the $p$-values are 0.047 and 0.078, respectively. These results indicate that the proposed attention enhanced pyramidal architecture improves prediction performance, and obtains higher correlations and lower prediction errors to other data-driven approaches. 

\begin{figure}[t]
	\centering
	\includegraphics[width=0.5\linewidth]{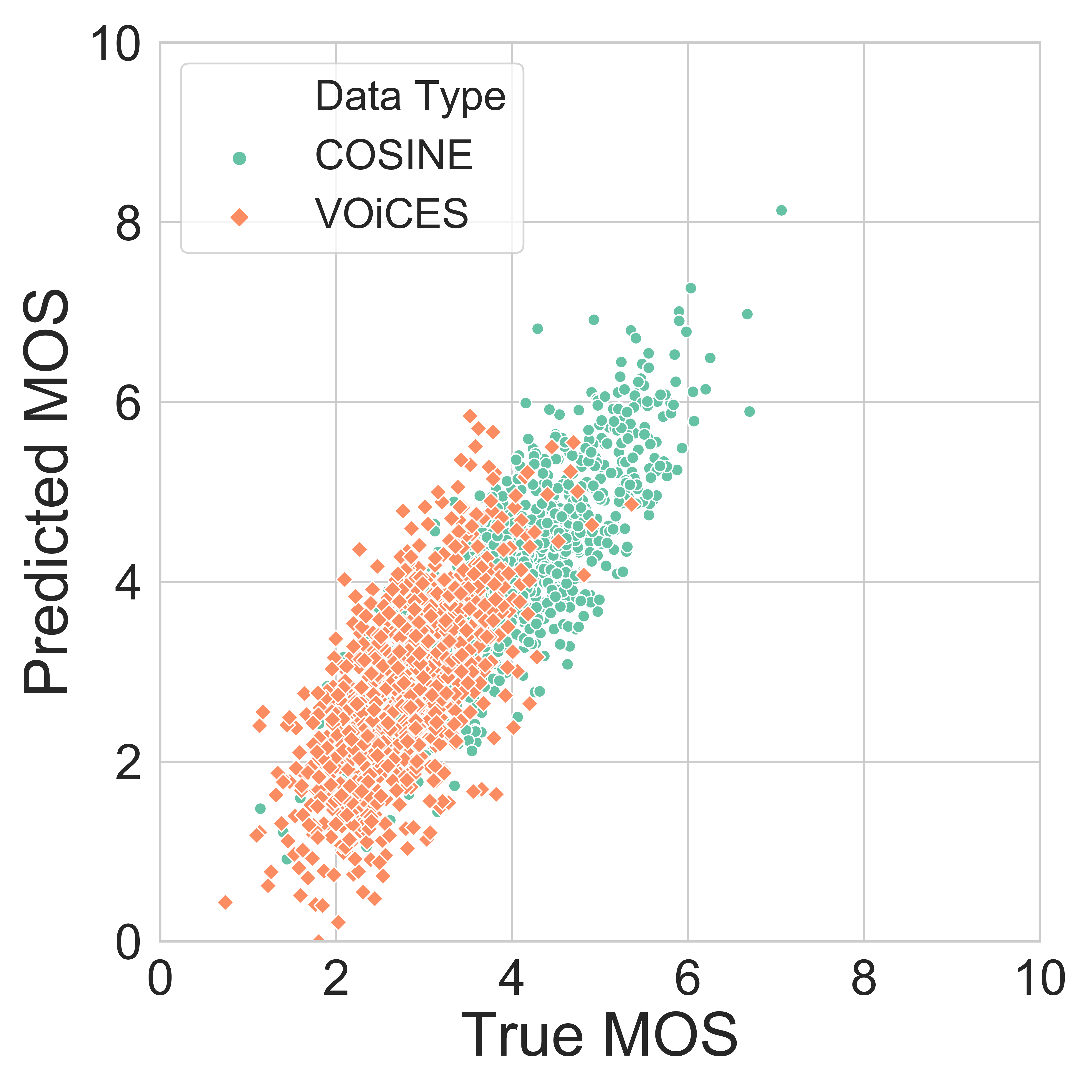}
	\caption{Correlation between the true MOS of the test stimuli and corresponding predicted MOS on COSINE (orange) and VOiCES (green) corpora.}\label{fig:mosscatterplot}
\end{figure}

Fig.~\ref{fig:mosscatterplot}, displays the relationship between the subjective MOS and the estimated MOS of the proposed approach. It can be seen that most predicted values scatter along the diagonal, which indicates high correlation with human MOS assessments.

\section{Conclusions}
\label{sec:conclusion}

In this paper, we present a data-driven approach to evaluate speech quality, by directly predicting human MOS ratings of real-world speech signals. A large-scale speech quality study is conducted using crowdsourcing to ensure that our prediction model performs accurately and robustly in real-world environments. An attention-based pyramid recurrent model is trained to estimate MOS. The experimental results demonstrate the superiority of the proposed model in contrast to the baseline models and several state-of-the-art methods in terms of speech quality evaluation. The collected dataset will also be made available to facilitate future research efforts.

\bibliographystyle{IEEEtran}

\bibliography{mybib}

\begin{thebibliography}{10}
\providecommand{\url}[1]{#1}
\csname url@samestyle\endcsname
\providecommand{\newblock}{\relax}
\providecommand{\bibinfo}[2]{#2}
\providecommand{\BIBentrySTDinterwordspacing}{\spaceskip=0pt\relax}
\providecommand{\BIBentryALTinterwordstretchfactor}{4}
\providecommand{\BIBentryALTinterwordspacing}{\spaceskip=\fontdimen2\font plus
\BIBentryALTinterwordstretchfactor\fontdimen3\font minus
  \fontdimen4\font\relax}
\providecommand{\BIBforeignlanguage}[2]{{%
\expandafter\ifx\csname l@#1\endcsname\relax
\typeout{** WARNING: IEEEtran.bst: No hyphenation pattern has been}%
\typeout{** loaded for the language `#1'. Using the pattern for}%
\typeout{** the default language instead.}%
\else
\language=\csname l@#1\endcsname
\fi
#2}}
\providecommand{\BIBdecl}{\relax}
\BIBdecl

\bibitem{hu2007evaluation}
Y.~Hu and P.~C. Loizou, ``Evaluation of objective quality measures for speech
  enhancement,'' \emph{IEEE Trans. Audio, Speech, Lang. Process.}, vol.~16,
  no.~1, 2007.

\bibitem{emiya2011subjective}
V.~Emiya, E.~Vincent, N.~Harlander, and V.~Hohmann, ``Subjective and objective
  quality assessment of audio source separation,'' \emph{IEEE Trans. Audio,
  Speech, Lang. Process.}, vol.~19, no.~7, 2011.

\bibitem{rec2008p}
ITU-T, ``P. 910: Subjective video quality assessment methods for multimedia
  applications,'' \emph{ITU Recommendation}, vol.~2, 2008.

\bibitem{series2014method}
ITU-R, ``{BS}. 1534： method for the subjective assessment of intermediate
  quality level of audio systems,'' \emph{ITU Recommendation}, 2014.

\bibitem{hirsch2000aurora}
H.-G. Hirsch and D.~Pearce, ``The aurora experimental framework for the
  performance evaluation of speech recognition systems under noisy
  conditions,'' in \emph{Automatic Speech Recognition: Challenges for the new
  Millenium ISCA Tutorial and Research Workshop (ITRW)}, 2000.

\bibitem{kinoshita2016summary}
K.~Kinoshita, M.~Delcroix, S.~Gannot \emph{et~al.}, ``A summary of the reverb
  challenge: state-of-the-art and remaining challenges in reverberant speech
  processing research,'' \emph{Journal on Advances in Signal Processing}, vol.
  2016, no.~1, 2016.

\bibitem{mclaren2016speakers}
M.~McLaren, A.~Lawson, L.~Ferrer, D.~Castan, and M.~Graciarena, ``The speakers
  in the wild speaker recognition challenge plan,'' \emph{Interspeech Special
  Session}, 2016.

\bibitem{reddy2020interspeech}
C.~K. Reddy, E.~Beyrami, H.~Dubey \emph{et~al.}, ``The interspeech 2020 deep
  noise suppression challenge: Datasets, subjective speech quality and testing
  framework,'' \emph{Interspeech Special Session}, 2020.

\bibitem{barker2018fifth}
J.~Barker, S.~Watanabe, E.~Vincent, and J.~Trmal, ``The fifth'chime'speech
  separation and recognition challenge: dataset, task and baselines,''
  \emph{arXiv preprint arXiv:1803.10609}, 2018.

\bibitem{pesq2001}
ITU-T, ``P. 862: Perceptual evaluation of speech quality ({PESQ}), an objective
  method for end-to-end speech quality assessment of narrowband telephone
  networks and speech codecs,'' \emph{ITU Recommendation}, 2001.

\bibitem{beerends2013perceptual}
J.~G. Beerends, C.~Schmidmer, J.~Berger \emph{et~al.}, ``Perceptual objective
  listening quality assessment ({POLQA}), the third generation {ITU-T} standard
  for end-to-end speech quality measurement part i—temporal alignment,''
  \emph{Journal of the Audio Engineering Society}, vol.~61, no.~6, pp.
  366--384, 2013.

\bibitem{cano2016evaluation}
E.~Cano, D.~FitzGerald, and K.~Brandenburg, ``Evaluation of quality of sound
  source separation algorithms: Human perception vs quantitative metrics,'' in
  \emph{Proc. EUSIPCO}.\hskip 1em plus 0.5em minus 0.4em\relax IEEE, 2016.

\bibitem{santos2014improved}
J.~F. Santos, M.~Senoussaoui, and T.~H. Falk, ``An improved non-intrusive
  intelligibility metric for noisy and reverberant speech,'' in \emph{Proc.
  IWAENC}.\hskip 1em plus 0.5em minus 0.4em\relax IEEE, 2014, pp. 55--59.

\bibitem{malfait2006p}
L.~Malfait, J.~Berger, and M.~Kastner, ``P. 563 - {T}he {ITU-T} standard for
  single-ended speech quality assessment,'' \emph{IEEE Trans. Audio, Speech,
  Lang. Process.}, vol.~14, no.~6, 2006.

\bibitem{kim2007anique+}
D.-S. Kim and A.~Tarraf, ``{ANIQUE}+: A new american national standard for
  non-intrusive estimation of narrowband speech quality,'' \emph{Bell Labs
  Technical Journal}, vol.~12, no.~1, 2007.

\bibitem{falk2010non}
T.~H. Falk, C.~Zheng, and W.-Y. Chan, ``A non-intrusive quality and
  intelligibility measure of reverberant and dereverberated speech,''
  \emph{IEEE Trans. Audio, Speech, Lang. Process.}, vol.~18, no.~7, pp.
  1766--1774, 2010.

\bibitem{mittag2019non}
G.~Mittag and S.~M{\"o}ller, ``Non-intrusive speech quality assessment for
  super-wideband speech communication networks,'' in \emph{Proc. ICASSP}.\hskip
  1em plus 0.5em minus 0.4em\relax IEEE, 2019, pp. 7125--7129.

\bibitem{fu2018quality}
S.-W. Fu, Y.~Tsao, H.-T. Hwang, and H.-M. Wang, ``Quality-{N}et: An end-to-end
  non-intrusive speech quality assessment model based on blstm,'' \emph{Proc.
  Interspeech}, 2018.

\bibitem{avila2019non}
A.~Avila, H.~Gamper, C.~Reddy, R.~Cutler \emph{et~al.}, ``Non-intrusive speech
  quality assessment using neural networks,'' in \emph{Proc. ICASSP}.\hskip 1em
  plus 0.5em minus 0.4em\relax IEEE, 2019, pp. 631--635.

\bibitem{dong2019classification}
X.~Dong and D.~S. Williamson, ``A classification-aided framework for
  non-intrusive speech quality assessment,'' in \emph{Proc. WASPAA}.\hskip 1em
  plus 0.5em minus 0.4em\relax IEEE, 2019, pp. 100--104.

\bibitem{dong2020attention}
------, ``An attention enhanced multi-task model for objective speech
  assessment in real-world environments,'' in \emph{Proc. ICASSP}.\hskip 1em
  plus 0.5em minus 0.4em\relax IEEE, 2020, pp. 911--915.

\bibitem{sharma2016data}
D.~Sharma, Y.~Wang, P.~A. Naylor, and M.~Brookes, ``A data-driven non-intrusive
  measure of speech quality and intelligibility,'' \emph{Speech Communication},
  vol.~80, pp. 84--94, 2016.

\bibitem{rec2006p}
ITU-T, ``P. 800.1: Mean opinion score ({MOS}) terminology,'' \emph{ITU
  Recommendation}, 2006.

\bibitem{paolacci2010running}
G.~Paolacci, J.~Chandler, and P.~G. Ipeirotis, ``Running experiments on amazon
  mechanical turk,'' \emph{Judgment and Decision making}, vol.~5, no.~5, pp.
  411--419, 2010.

\bibitem{stupakov2009cosine}
A.~Stupakov, E.~Hanusa, J.~Bilmes \emph{et~al.}, ``{COSINE}-a corpus of
  multi-party conversational speech in noisy environments,'' in \emph{Proc.
  ICASSP}.\hskip 1em plus 0.5em minus 0.4em\relax IEEE, 2009, pp. 4153--4156.

\bibitem{richey2018voices}
C.~Richey, M.~Barrios, Z.~Armstrong \emph{et~al.}, ``Voices obscured in complex
  environmental settings ({VOICES}) corpus,'' \emph{arXiv preprint
  arXiv:1804.05053}, 2018.

\bibitem{ribeiro2011crowdmos}
F.~Ribeiro, D.~Flor{\^e}ncio, C.~Zhang, and M.~Seltzer, ``Crowd{MOS}: An
  approach for crowdsourcing mean opinion score studies,'' in \emph{Proc.
  ICASSP}.\hskip 1em plus 0.5em minus 0.4em\relax IEEE, 2011, pp. 2416--2419.

\bibitem{schoeffler2015towards}
M.~Schoeffler, F.-R. St{\"o}ter, B.~Edler, and J.~Herre, ``Towards the next
  generation of web-based experiments: A case study assessing basic audio
  quality following the {ITU-R} recommendation {BS}. 1534 ({MUSHRA}),'' in
  \emph{1st Web Audio Conference}, 2015, pp. 1--6.

\bibitem{cartwright2016fast}
M.~Cartwright, B.~Pardo, G.~J. Mysore, and M.~Hoffman, ``Fast and easy
  crowdsourced perceptual audio evaluation,'' in \emph{Proc. ICASSP}.\hskip 1em
  plus 0.5em minus 0.4em\relax IEEE, 2016, pp. 619--623.

\bibitem{chorowski2015attention}
J.~K. Chorowski, D.~Bahdanau, D.~Serdyuk, K.~Cho, and Y.~Bengio,
  ``Attention-based models for speech recognition,'' in \emph{NIPS}, 2015, pp.
  577--585.

\bibitem{chan2016listen}
W.~Chan, N.~Jaitly, Q.~Le, and O.~Vinyals, ``Listen, attend and spell: A neural
  network for large vocabulary conversational speech recognition,'' in
  \emph{Proc. ICASSP}.\hskip 1em plus 0.5em minus 0.4em\relax IEEE, 2016.

\bibitem{rec2018crowdspeech}
ITU-T, ``P. 808: Subjective evaluation of speech quality with a crowdsourcing
  approach,'' \emph{ITU Recommendation}, 2018.

\bibitem{naderi2015effect}
B.~Naderi, T.~Polzehl, I.~Wechsung, F.~K{\"o}ster, and S.~M{\"o}ller, ``Effect
  of trapping questions on the reliability of speech quality judgments in a
  crowdsourcing paradigm,'' in \emph{Conference of the International Speech
  Communication Association}, 2015.

\bibitem{gadiraju2015understanding}
U.~Gadiraju, R.~Kawase, S.~Dietze, and G.~Demartini, ``Understanding malicious
  behavior in crowdsourcing platforms: The case of online surveys,'' in
  \emph{ACM Conference on Human Factors in Computing Systems}, 2015, pp.
  1631--1640.

\bibitem{zielinski2008some}
S.~Zielinski, F.~Rumsey, and S.~Bech, ``On some biases encountered in modern
  audio quality listening tests-a review,'' \emph{Journal of the Audio
  Engineering Society}, vol.~56, no.~6, pp. 427--451, 2008.

\bibitem{han2011data}
J.~Han, J.~Pei, and M.~Kamber, \emph{Data mining: concepts and
  techniques}.\hskip 1em plus 0.5em minus 0.4em\relax Elsevier, 2011.

\bibitem{ester1996density}
M.~Ester, H.-P. Kriegel, J.~Sander, X.~Xu \emph{et~al.}, ``A density-based
  algorithm for discovering clusters in large spatial databases with noise.''
  in \emph{Proc. KDD}, vol.~96, no.~34, 1996, pp. 226--231.

\bibitem{liu2008isolation}
F.~T. Liu, K.~M. Ting, and Z.-H. Zhou, ``Isolation forest,'' in \emph{Proc.
  ICDM}.\hskip 1em plus 0.5em minus 0.4em\relax IEEE, 2008, pp. 413--422.

\bibitem{zhang2019sequence}
J.-X. Zhang, Z.-H. Ling, L.-J. Liu, Y.~Jiang, and L.-R. Dai,
  ``Sequence-to-sequence acoustic modeling for voice conversion,'' \emph{IEEE
  Trans. Audio, Speech, Lang. Process.}, vol.~27, no.~3, pp. 631--644, 2019.

\bibitem{vaswani2017attention}
A.~Vaswani, N.~Shazeer, N.~Parmar \emph{et~al.}, ``Attention is all you need,''
  in \emph{NIPS}, 2017, pp. 5998--6008.

\bibitem{rec2012p}
ITU-T, ``P. 1401, methods, metrics and procedures for statistical evaluation,
  qualification and comparison of objective quality prediction models,''
  \emph{ITU Recommendation}, 2012.

\bibitem{patton2016automos}
B.~Patton, Y.~Agiomyrgiannakis, M.~Terry, K.~Wilson, R.~A. Saurous, and
  D.~Sculley, ``Auto{MOS}: Learning a non-intrusive assessor of
  naturalness-of-speech,'' \emph{NIPS Workshop}, 2016.

\end{thebibliography}

\end{document}